\documentclass[a4paper]{INTERSPEECH2023}
\interspeechcameraready 
\newcommand{\Fig}[1]{\textbf{Fig. \ref{fig:#1}}} 
\newcommand{\Figs}[1]{\textbf{Figs. \ref{fig:#1}}} 

\newcommand{\Eq}[1]{Eq. (\ref{eq:#1})} 
\newcommand{\Eqs}[1]{Eqs. (\ref{eq:#1})}
\newcommand{\Eqss}[1]{(\ref{eq:#1})}

\newcommand{\Norm}[1]{ {\mathcal N}\left( #1 \right) } 
 
\renewcommand{\Vec}[1]{\textrm{\boldmath $#1$}} 

\newcommand{\pt}[1]{\left(#1\right)} 
\newcommand{\x}{ \Vec{x} } 
\newcommand{\z}{ \Vec{z} } 
\newcommand{\rr}{ \Vec{r} } 
\newcommand{\hatx}{ \Vec{\hat x} } 

\newcommand{\drawfig}[4]{ 
  \begin{figure}[#1]
  \centering \vspace{-0mm}
  \includegraphics[width=#2,clip]{#3.pdf} \vspace{-3mm} 
  \caption{#4} \vspace{-5mm}
  \label{fig:#3}
  \end{figure}
}
\usepackage{multirow}
\title{HumanDiffusion: diffusion model using perceptual gradients}
\name{Yota Ueda, Shinnosuke Takamichi, Yuki Saito, \\Norihiro Takamune, Hiroshi Saruwatari}
\address{
  The University of Tokyo, Japan.}
\email{weda@g.ecc.u-tokyo.ac.jp, shinnosuke\_takamichi@ipc.i.u-tokyo.ac.jp}

\begin{document}
\maketitle

\setlength{\abovedisplayskip}{3pt}
\setlength{\belowdisplayskip}{3pt}

    \vspace{-2mm}
\begin{abstract}
    \vspace{-1mm}
    We propose {\it HumanDiffusion,} a diffusion model trained from humans' perceptual gradients to learn an acceptable range of data for humans (i.e., human-acceptable distribution). Conventional HumanGAN aims to model the human-acceptable distribution wider than the real-data distribution by training a neural network-based generator with human-based discriminators. However, HumanGAN training tends to converge in a meaningless distribution due to the gradient vanishing or mode collapse and requires careful heuristics. In contrast, our HumanDiffusion learns the human-acceptable distribution through Langevin dynamics based on gradients of human perceptual evaluations. Our training iterates a process to diffuse real data to cover a wider human-acceptable distribution and can avoid the issues in the HumanGAN training. The evaluation results demonstrate that our HumanDiffusion can successfully represent the human-acceptable distribution without any heuristics for the training.
\end{abstract}

\noindent\textbf{Index Terms}:  diffusion models, human computation, black-box optimization, crowdsourcing, speech perception

\vspace{-2mm}
\section{Introduction}
\vspace{-2mm}
\label{sec:introduction}
Generative models can produce data that are indistinguishable from real data. In particular, deep generative models~\cite{goodfellow2020generative,kingma2013vae,dinh2014nice,rombach2022high}, which are based on deep neural networks (DNNs), have significantly improved the quality of generated data in media research, including speech synthesis~\cite{saito18advss, li2021starganv2, casanova2022yourtts}, natural language processing~\cite{floridi2020gpt}, and image synthesis~\cite{choi2020stargan}. The generative models can represent real-data distributions and produce data that follow the learned distributions~\cite{goodfellow2020generative}. In contrast, humans may accept data as natural even when the data are outliers of a real-data distribution~\cite{fujii2020humangan}. For example, in speech perception, humans can accept synthesized or processed speech. In this paper, we use the term \textit{human-acceptable distribution} defined as the data distribution whose data humans can accept as natural~\cite{fujii2020humangan}. 

HumanGAN~\cite{fujii2020humangan} was proposed to model a human-acceptable distribution, whereas the basic generative adversarial network (GAN)~\cite{goodfellow2020generative} can only represent a real-data distribution. GAN is a type of deep generative model and consists of a DNN-based generator and a DNN-based discriminator. HumanGAN replaces the discriminator of GAN with a human-based discriminator. HumanGAN regards humans as black-box systems that output perceptual evaluation values, given the generated data. By using the estimated perceptual gradient of the human-based discriminator, one can then train a generator to represent a human-acceptable distribution. However, HumanGAN often suffers from gradients vanishing and mode collapse during training similar to the original GAN~\cite{zhang2018convergence}. In other words, it requires careful heuristics to prevent the learned distribution from converging in a meaningless region, such as outside the human-acceptable distribution or around data whose evaluation value is extremely high (i.e., the mode of perceptual evaluation).

To solve these problems, we propose \textit{HumanDiffusion}, a diffusion model using perceptual gradients. For modeling a human-acceptable distribution, we introduce diffusion models~\cite{song2020score} to sample data that follows a real-data distribution using the gradient of the distribution.  \Fig{gan_diff} shows a comparison of HumanDiffusion and HumanGAN, and the distributions these models represent. On the basis of a diffusion model, our HumanDiffusion iterates a process to diffuse real data to cover a wider human-acceptable distribution thereby avoiding the issues encountered during the HumanGAN training. HumanDiffusion is evaluated in terms of its phoneme perception. The experimental results show that HumanDiffusion can successfully represent a human-acceptable distribution.

\drawfig{t}{0.99\linewidth}{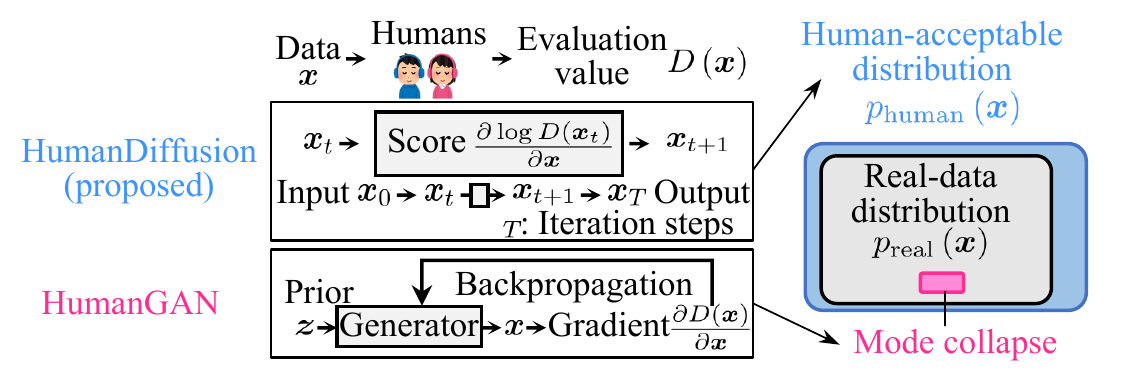}{HumanDiffusion and HumanGAN. HumanDiffusion can represent a human-acceptable distribution whereas HumanGAN cannot without a heuristic solution.}

\vspace{-2mm}
\section{Related Work}
\vspace{-2mm}
Human perception is incorporated into DNNs such as reinforcement learning~\cite{lambert2022illustrating}, and genetic algorithm~\cite{asadi2015human}. HumanGAN~\cite{fujii2020humangan} deals with perceptual gradients to train a generator. Because humans are better at relative evaluation than absolute evaluation, gradient-based methods can train DNNs well.
    
    \vspace{-1mm}
    \subsection{HumanGAN}
    \vspace{-2mm}
    \label{subsec:Humangan}
        HumanGAN~\cite{fujii2020humangan} represents the human-acceptable distribution $p_\mathrm{human}\pt{\x}$, where $\x$ is data, that is wider than the real-data distribution $p_\mathrm{real}\pt{\x}$.
        A DNN-based generator of HumanGAN is trained using a human-based discriminator (humans' perceptual evaluation) instead of the DNN-based discriminator of GAN~\cite{goodfellow2020generative}. Let $N$ be the number of data. The generator $G\pt{\cdot}$ transforms prior data $\{\z_n\}_{1\leqq n \leqq N}$ produced from a prior distribution $\pi\pt{\z}$ to data $\{\hatx_n\}_{1\leqq n \leqq N}$. The human-based discriminator $D\pt{\cdot}$ imitates humans' perceptual evaluations. $D\pt{\cdot}$ takes $\hatx_n$ as an input and outputs a posterior probability that the input is perceptually acceptable. The objective function is
            \begin{align}
                L_\mathrm{HumanGAN} = \sum\limits_{n=1}^N D\pt{G\pt{\z_n}}.
                \label{eq:humangan_loss}
            \end{align}
        $G\pt{\cdot}$ is trained to maximize $L_\mathrm{HumanGAN}$. A model parameter $\bm{\theta}$ of $G\pt{\cdot}$ is iteratively updated as $\bm\theta \leftarrow \bm\theta + \alpha \partial L_\mathrm{HumanGAN}/{\partial \bm\theta}$, where $\alpha$ is the learning rate, and $\partial L_\mathrm{HumanGAN} / \partial \bm\theta = \partial L_\mathrm{HumanGAN} / \partial \x \cdot \partial \x / \partial \bm\theta
        = \partial D\pt{\x} / \partial \x \cdot \partial \x / \partial \bm\theta$.
        
        $\partial\x/\partial\bm\theta$ can be estimated analytically, but $\partial D\pt{\x}/{\partial \x}$ cannot because the human-based discriminator $D\pt{\cdot}$ is not differentiable.
        HumanGAN uses the natural evolution strategy (NES)~\cite{ilyas18blackboxlimitedquery} algorithm to approximate the gradient. A small perturbation $\Delta \x_{n}^{(i)}$ is randomly generated from the multivariate Gaussian distribution $\Norm{\Vec{0}, \sigma_\mathrm{NES}^2\Vec{I}}$, and it is added to the generated data $\hatx_n$. 
        $i$ is the perturbation index $\pt{1 \leq i \leq I}$. $\sigma_\mathrm{NES}$ and $\Vec{I}$ are the standard deviation and the identity matrix, respectively.
        Next, a human observes two perturbed data $\{ \hatx_n + \Delta \x_{n}^{(i)}, \hatx_n - \Delta \x_{n}^{(i)} \}$ and evaluates their difference in the posterior probability of naturalness:
            \begin{align}
                \Delta D\pt{\hatx_n^{(i)}} \equiv D\pt{\hatx_n + \Delta \x_{n}^{(i)}}- D\pt{\hatx_n - \Delta \x_{n}^{(i)}}.
                \label{eq:D}
            \end{align}
        $\Delta D\pt{\hatx_n^{(i)}}$ ranges from $-1$ to $1$. For instance, a human will answer $\Delta D\pt{\hatx_n^{(i)}}=1$ when the human perceives that $\hatx_n + \Delta \x_{n}^{(i)}$ is substantially more acceptable than $\hatx_n - \Delta \x_{n}^{(i)}$. $\partial D\pt{\hatx_n} / \partial \x$ is approximated as~\cite{ilyas18blackboxlimitedquery}
            \begin{align}
                \frac{\partial D\pt{\hatx_n}}{\partial \x}&= \frac{1}{2\sigma_\mathrm{NES} I} \sum\limits_{i=1}^{I} 
                \Delta D\pt{\hatx_n^{(i)}} \cdot \Delta \x_{n}^{(i)}. 
                \label{eq:grad}
            \end{align}

    However, HumanGAN requires various heuristics during training, such as initialization and training stops.
    First, HumanGAN should roughly know the shapes of the human-acceptable distribution in advance to make the training easier. Data that deviate significantly from the human-acceptable distribution cannot be distinguished by humans, resulting in an inaccurate estimation of the gradient in \Eq{grad}. Moreover, the generator training should be stopped before converging to represent the human-acceptable distribution because the training causes mode collapse due to the training based on the maximization of $L_{\rm human}$. Concretely, if the generator finds one region in the data space where humans evaluate them as highly acceptable during the training, the learned distribution converged in a limited range of the whole human-acceptable distribution.
    \vspace{-1mm}
    
    \subsection{Diffusion models}
    \vspace{-1mm}
    The diffusion models~\cite{song2020score} use the gradient of the data distribution to train a model that represents the real-data distribution $p_\mathrm{real}\pt{\x}$. 
    The score of the real-data distribution $p_\mathrm{real}\pt{\x}$ is defined as $\frac{\partial\log p_\mathrm{real}\pt{\x}}{\partial \x}$. The score network $S\pt{\cdot}$ is parameterized by $\bm{\theta}$ and trained to approximate the score. 
    
    Even if the true $p_\mathrm{real}\pt{\x}$ is not observable, as long as the score $\frac{\partial\log p_\mathrm{real}\pt{\x}}{\partial \x}$ is observable, an iterative update with Langevin dynamics~\cite{welling2011bayesian} can be applied as follows. Langevin dynamics is a kind of Markov chain Monte Carlo (MCMC) method using the gradients of the data distributions.
    First, an input data $\x_0$ is sampled from a prior distribution $\pi\pt{\z}$. Then, the data at $(t+1)$th step ($t = 0, 1, \ldots, T-1$) is calculated based on the Langevin dynamics using the score:
     \begin{align}
     \x_{t+1}=\x_t+\frac{\epsilon^2}{2} \frac{\partial \log p_\mathrm{real}\left(\x_t\right)}{\partial \x}+\epsilon \bm{r},
     \label{eq:langevin}
     \end{align}
    where $\epsilon > 0$ and $\bm{r}$ are a fixed step size and random vector from the isotropic Gaussian $N\pt{\bm{0}, \bm{I}}$, respectively. Finally, the distribution of the output data at the final step $T$, $\x_T$, will equal $p_\mathrm{real}\pt{\x}$ when $\epsilon \rightarrow 0$ and $T \rightarrow \infty$. Because sampling from \Eq{langevin} only requires the score $\frac{\partial\log p_\mathrm{real}\pt{\x}}{\partial \x}$, the score network $S\pt{\x} \approx \frac{\partial\log p_\mathrm{real}\pt{\x}}{\partial \x}$ can be trained, and it can then approximately produce samples with Langevin dynamics. Given real data $\{\x_n\}_{1\leqq n \leqq N}$, the score $\frac{\partial \log p_\mathrm{real}\pt{\x}}{\partial \x}$ is predicted by using $S\pt{\cdot}$. 
    
    The score network $S\pt{\x;\bm{\theta}}$ for estimating $\frac{\partial \log p_\mathrm{real}\pt{\x}}{\partial \x}$ is trained by score matching without training a model to estimate $p_\mathrm{real}\pt{\x}$. The objective function for training $S\pt{\cdot}$ is
     \begin{align}
     L_{\mathrm{Diffusion}} = \sum\limits_{n=1}^N \left|S\pt{\x_n;\bm{\theta}} - \frac{\partial \log p_\mathrm{real}\pt{\x_n}}{\partial \x}\right|^2.
     \label{eq:diffusion_loss}
     \end{align}
     
    As explained above, the diffusion model can generate data that follow a real-data distribution by using scores of the real-data distribution. If we can observe scores of the human-acceptable distribution instead of the real-data distribution, and if the score network can learn them, we expect to be able to generate data according to the human-acceptable distribution. 

\vspace{-2mm}
\section{HumanDiffusion}
\vspace{-1mm}
\label{sec:humandiff}
\subsection{Components of HumanDiffusion}
    \vspace{-1mm}
    \drawfig{t}{0.99\linewidth}{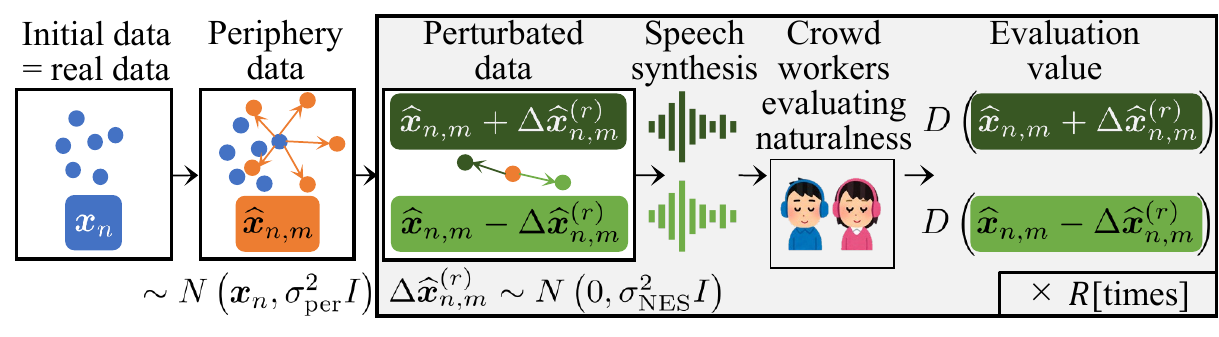}
    {Tasks for crowdworkers to estimate the score. Periphery data are sampled from real data. A human observes two perturbed data and evaluates their perceptual difference in naturalness acceptability. The score is estimated using evaluation and perturbation.}

    \drawfig{t}{0.99\linewidth}{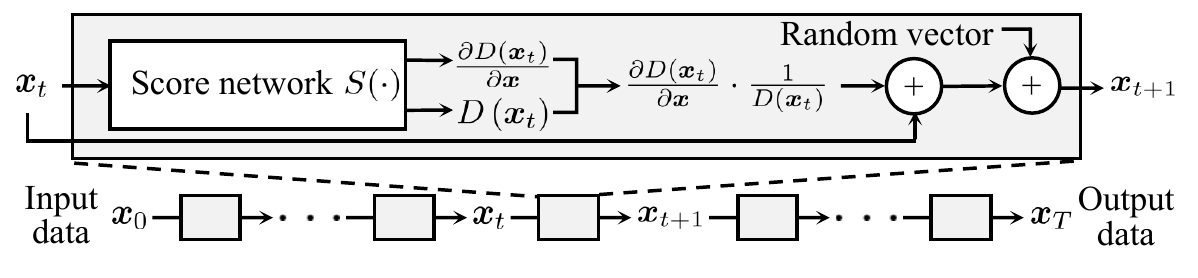}
    {Training score network and inference in HumanDiffusion. The score network outputs gradients and evaluation values to calculate scores. Samples are produced with Langevin dynamics.}
    
    \Fig{crowdwork} shows the process of  HumanDiffusion. HumanDiffusion uses the gradient of perceptual evaluation to train a model representing the human-acceptable distribution $p_{\mathrm{human}}\pt{\x} = \frac{1}{Z} D\pt{\x}$, where $Z = \int D\pt{\x} d\x$ is the normalization coefficient.
    The score is defined as $\frac{\partial\log D\pt{\x}}{\partial \x}$. We process \textit{periphery data} sampled from real data $\x_n$ to estimate the score around the real-data distribution. Let $M$ be the number of periphery data per real data. We sample periphery data $\{\hatx_{n,m}\}_{1\leqq n \leqq N, 1\leqq n \leqq M}$ from $N(\x_n, \sigma_{\mathrm{per}}^2\bm I)$, where $\sigma_\mathrm{per}$ is the standard deviation. The perceptual evaluation function $D\pt{\cdot}$ is driven by humans and takes data $\x$ as the input and outputs the perceptual evaluation value of the naturalness of the data in the range of $[0, 1]$. If the data is natural, the evaluation value is $1$, and if the data is unnatural, the evaluation value is $0$. The score network $S\pt{\cdot}$ is trained to approximate the score of $D\pt{\x}$. 
    The objective function is defined as the minimization of 
        \begin{align}
            \label{eq:loss_humandiff}
            L_\mathrm{HumanDiff} = \sum\limits_{n=1}^N\sum\limits_{m=1}^M \left|S\pt{\x_{n,m};\bm{\theta}} - \frac{\partial\log D \pt{\hatx_{n,m}}}{\partial\x}\right|^2.
        \end{align}
    
    \subsection{Estimating score of perceptual evaluation}
    \vspace{-1mm}
        \label{sub:direct}
        Similar to the HumanGAN, the score $\frac{\partial \log D\pt{\x}}{{\partial \x}}$ cannot be estimated because the evaluation functions are based on the perceptual evaluation. We use the NES algorithm to estimate the score. The NES algorithm uses evaluation values of two perturbed data. With the index $i$, a human observes two perturbed data $\hatx_{n,m}, + \Delta \x_{n,m}^{(i)}, \hatx_{n,m} - \Delta \x_{n,m}^{(i)}$, where $\Delta\x_{n,m}$ is sampled from $N(0, \sigma_{\mathrm{NES}}^2\bm I)$ in the same way as \Eq{grad}. In HumanGAN, humans responded only to the difference between the two samples (\Eq{D}), but in HumanDiffusion, humans responded to the absolute value of naturalness for each of these two samples.  

    \label{subsubsec:gradient}
        
        To estimate the score using the NES algorithm in the same manner as using HumanGAN, we adopt the chain rule to the score: $\frac{\partial \log D\pt{\x}}{\partial \x}=\frac{1}{D\pt{\x}}\frac{\partial D\pt{\x}}{\partial \x}$. The score is obtained by estimating $D\pt{\x}$ and $\frac{\partial D\pt{\x}}{\partial\x}$. We remodel the score network $S\pt{\cdot}$ to output $S_D\pt{\cdot}\approx D\pt{\x}$ and $S_{\nabla D}\pt{\cdot}\approx \frac{\partial D\pt{\hatx_{n,m}}}{\partial \x}$ instead of $\frac{\partial\log D\pt{\hatx_{n,m}}}{\partial \x}$, as shown in \Fig{scorenet}. If the dimension of the data is $d$, this network outputs values in $d+1$ dimensions.
        The loss function of the score network (\Eq{loss_humandiff}) is redefined as
        \begin{align}
        L^\mathrm{unconditional}_\mathrm{HumanDiff} =& \sum\limits_{n=1}^N\sum\limits_{m=1}^M \bigl(\left|S_D\pt{\hatx_{n,m}} - D\pt{\hatx_{n,m}}\right|^2
        \nonumber\\
        &+\left|S_{\nabla D}\pt{\hatx_{n,m}} - \frac{\partial\log D\pt{\hatx_{n,m}}}{\partial \x} \right|^2\bigr).
        \end{align}
        $\frac{\partial D\pt{\x}}{\partial\x}$ can be approximated in the same manner as in HumanGAN:
        \begin{align}
        \label{eq:gradient}
            \frac{\partial D\pt{\hatx_{n,m}}}{\partial \x}=&
            \frac{1}{2\sigma_\mathrm{NES} I} 
            \sum\limits_{i=1}^{I} \biggl\{ D\pt{\hatx_{n,m}+\Delta \x^{(i)}_{n,m}}
            \nonumber\\
            &- D\pt{\hatx_{n,m}-\Delta \x^{(i)}_{n,m}}\biggr\}\cdot \Delta \x_{n,m}^{(i)}.
        \end{align}
        
        Next is the estimation of $D\pt{\x}$, where $D^{(i)}\pt{\hatx_{n,m}}$ at the index $i$ is given by 
            $D^{(i)}(\hatx_{n,m})=\pt{D\pt{\hatx_{n,m}+\Delta \x_{n,m}^{(i)}}
            \nonumber\\
            +D\pt{\hatx_{n,m}-\Delta \x_{n,m}^{(i)}}}/2$, and $D\pt{\hatx_{n,m}}$ is estimated from it. The most intuitive method is to estimate the mean $D\pt{\hatx_{n,m}}=\frac{1}{I} \sum\limits_{i=1}^I D^{(i)}\pt{\hatx_{n,m}}$. However, since $D^{(i)}\pt{\hatx_{n,m}}$ is a defined range of values, the distribution of $D^{(i)}\pt{\x_{n,m}}$ is not symmetric near its lower bound (0, i.e., the perceived quality of the data is extremely poor) or, its upper bound (1, the perceived quality is extremely good). Therefore, the mean is not a suitable representative value of the distribution. Therefore, in this paper, we use kernel distribution estimation given $P\pt{D^{(i)}\pt{\hatx_{n,m}}}$ and consider its mode as $D\pt{\hatx_{n,m}}$. Preliminary experiments were conducted using both the mean and mode methods, and the mean did not show the sampling convergence.

        From the above formulation, the score network is a network that outputs both $D\pt{\x}$ and $\frac{\partial D\pt{\x}}{\partial\x}$, as shown in \Fig{scorenet}. If the dimension of the data is $d$, this network outputs values in $d+1$ dimensions. The score network is trained using \Eq{loss_humandiff} by computing $\frac{\partial\log D\pt{\x}}{\partial\x}$ from the output $D\pt{\x}$ and $\frac{\partial D\pt{\x}}{\partial\x}$. 
        
     
    \subsubsection{Compensating for inaccurate gradient}
        If two perturbated data of extremely poor quality (e.g., sounds that do not sound like a speech) are observed and evaluated by a human, the score is very inaccurate\footnote{Intuitively, it is difficult to evaluate the naturalness of two extremely poor quality data. Therefore, the score of periphery data $\x_{n,m}$ of extremely poor quality cannot be estimated from the gradient alone.}. Therefore, we assume that the naturalness of the real data is not poor at all, and we apply a regularization that brings the gradient $\frac{\partial D\pt{\x}}{\partial \x}$ closer to that of the real data used to sample the periphery data as
        \begin{align}
        \label{eq:compensate}
        \frac{\partial D\pt{\hatx_{n,m}}}{\partial \x} \leftarrow \frac{\partial D\pt{\hatx_{n,m}}}{\partial \x}+b\pt{\hatx_{n,m}-\x_n},
        \end{align}
        where $b$ is a hyperparameter. 
        This method is based on an existing diffusion model that robustly estimates scores in the absence of any data~\cite{song2020score}. Preliminary experiments have confirmed that without this regularization, the scores for extremely poor-quality data are very close to zero, and the sampling described below does not converge.
        
    \subsection{Sampling with Langevin dynamics}
        The score is computed using \Eqs{gradient} and \Eqss{compensate} and modeled them using \Eq{loss_humandiff}. Langevin dynamics is used to sample the data according to the human-acceptable distribution $p_\mathrm{human}\pt{\x}$.
        \begin{align}
        \x_{t+1}=\x_t+\frac{\epsilon^2}{2}  S\left(\x_t\right)+\epsilon \rr.
        \label{langevin}
        \end{align}
        The initial distribution that $\x_0$ follows is a real-data distribution (e.g., data generated by a trained GAN). The training iteration is continued until convergence.

\vspace{-2mm}
\section{Experiments and Results}
    \vspace{-1mm}
    \subsection{Experimental setup} \label{sec:expcond}
    \vspace{-1mm}
        \textbf{Data space and speech analysis-synthesis.}
        Phoneme perception experiments were conducted as described in a previous report on HumanGAN using basically the same experimental setup~\cite{fujii2020humangan}. We basically followed the experimental setup of the HumanGAN paper~\cite{fujii2020humangan}, specifically, we used the phoneme (Japanese /a/), the JVPD corpus~\cite{jvpd_corpus}, preprocessing, the WORLD vocoder~\cite{morise16world,morise16d4c}, and speech features including log spectral envelopes. We applied principal component analysis (PCA) to the log spectral envelopes and used the first and second principal components. The two-dimensional principal components were normalized to have zero-mean and unit-variance. The speech synthesis step also followed the previous report on HumanGAN paper~\cite{fujii2020humangan}. First, the first and second principal components were generated by the neural network and de-normalized. Then, a 1-second speech sample was synthesized using the remaining speech features, i.e., F0, and aperiodicities. The evaluation was performed by using the Lancers crowdsourcing platform~\cite{lancers}. 

        \textbf{Perceptual test.}
        We carried out perceptual evaluations. Two speech waveforms generated from $\hatx_{n,m}+\Delta \x_{n,m}^{(i)}$, $\hatx_{n,m}-\Delta \x_{n,m}^{(i)}$ were presented to a listener, and the listener provided the naturalness values $D\pt{\hatx_{n,m}+\Delta \x_{n,m}^{(i)}}$ and $D\pt{\hatx_{n,m}-\Delta \x_{n,m}^{(i)}}$ using slide bars. The slide bars range from $0$ to $1$. Listeners selected $1$ if the speech was  natural and $0$ when the speech was unnatural. The total number of listeners was $150$. 
        
        \textbf{Score network and HumanDiffusion sampling.}
        \label{scorenet}
        The score network was a feed-forward neural network. It inputs two-dimensional data $\x$ and outputs one-dimensional $D\pt{\x}$ and two-dimensional
        $\frac{\partial D\pt{\x}}{\partial \x}$. The network consisted of a two-unit input layer, $3\times128$-unit softplus~\cite{liu2016large} hidden layers, and three-unit sigmoid (for $D\pt{\x}$) and linear (for $\frac{\partial D\pt{\x}}{\partial \x}$) output layers. We used the Adam~\cite{kingma14adam} optimizer  with a learning rate of $\alpha = 0.001$ for the training. The number of real data $N$, the number of periphery data per real datum $M$, the number of perturbations $I$, the number of training iterations, and the standard deviation of NES $\sigma_\mathrm{NES}$, the standard deviation $\sigma_\mathrm{per}$ were set to $100$, $3$, $20$, $10000$, $1.0$ and $10$, respectively. For the Langevin dynamics sampling, the number of data sampled from real-data distribution $N\pt{0, 1}$, step size $\epsilon$, the number of iterations were set to $200$, $0.0001$, $100000$. We confirmed that the distribution converged after $10,000$ iterations of the sampling.

            \drawfig{t}{0.99\linewidth}{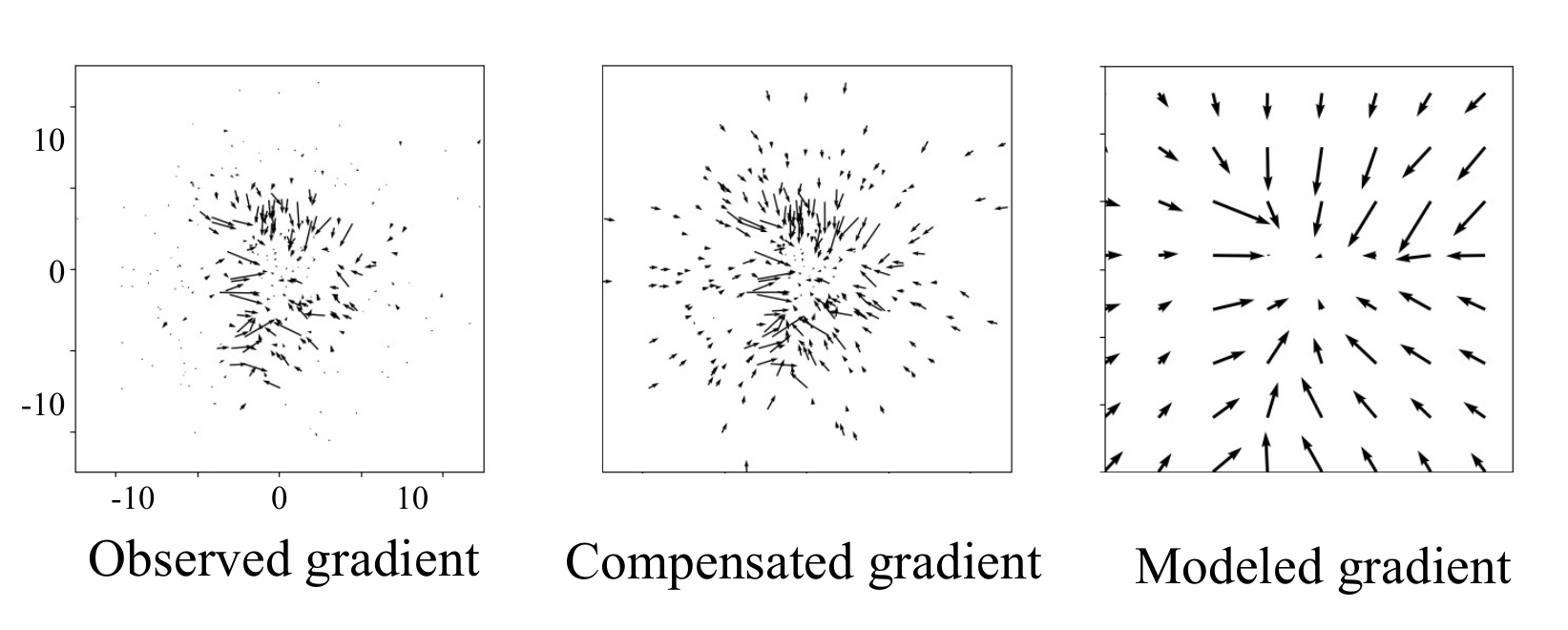}
    {Observed gradients, gradients with regularization, and modeled gradient by score network.}
        \textbf{HumanGAN.}
        For comparison, we also trained HumanGAN. The generator was a feed-forward neural network. The model consisted of a two-unit input layer, $3\times128$-unit softplus hidden layers, and a two-unit linear output layer. Although the original training of HumanGAN requires listening experiments to determine the perceptual gradient for each iteration, in this study, we used the output of the score network to reduce the cost of experiments. Specifically, the output $\frac{\partial D\pt{\x}}{\partial \x}$ of the score network was used for training. We used the Adam optimizer with a learning rate of $\alpha = 0.01$ for the training. The number of training iterations was set to $10000$.
        The initial parameters were set to cover a neighborhood of the perceptual distribution as described in the report on HumanGAN~\cite{fujii2020humangan}.
        
    \vspace{-1mm}
    \subsection{Qualitative evaluation of score network training}
    \vspace{-1mm}
        
        First, we qualitatively confirm the observed scores and the score network training. The left panel in \Fig{gradient} is the estimated gradient for each $\frac{\partial D\pt{\hatx_{n,m}}}{\partial \x}$. As described in Section \ref{eq:grad}, the real data distribution follows a mean of $0$ and a variance of $1$. This figure shows that a gradient occurs over a wider range than that of the real data distribution. In other words, as shown in the HumanGAN paper, the perceptual distribution exists over a wider range than the real data distribution. On the other hand, when the data are outliers far from the real data distribution (e.g., when the x-axis value exceeds $\pm 5$), the gradient is almost zero. This indicates that the perceptual gradient has vanished owing to extremely poor sound quality.
        Regularization compensates for this gradient vanishing, and the result is shown in the middle panel of \Fig{gradient}. This regularization produces a gradient toward the real data distribution even for data that are outliers. The result of training this gradient in the score network is shown on the right panel of the figure. Finally, a gradient toward the center is learned.
        
    \vspace{-1mm}
    \subsection{Quantitative/qualitative evaluation of sampling}
    \vspace{-1mm}
        
        Next, we compare the data generated by HumanGAN and HumanDiffusion to show that the HumanDiffusion sampling is reasonable. \Figs{train} a) and b) show the distributions of data generated by HumanGAN and HumanDiffusion, respectively. The gradients are also shown for reference.
        
        HumanGAN learns only to move data in the gradient direction, which causes mode collapse, as shown on the right panel of \Fig{train} a) unless a heuristic iteration limit is set. In contrast, HumanDiffusion includes a stochastic movement term based on MCMC and thus captures a wider range than the real data distribution, as shown on the right panel of \Fig{train} b). The results demonstrate that HumanDiffusion can learn a wider human-acceptable distribution without introducing heuristics used in HumanGAN.
        The variance of the real data distribution is $[1.0, 1.0]$ (unit variance), whereas the variance of the data distribution obtained by HumanDiffusion is $[9.2, 8.9]$, indicating that HumanDiffusion represents a wider distribution than the real data distribution. 
        

    \drawfig{t}{0.9\linewidth}{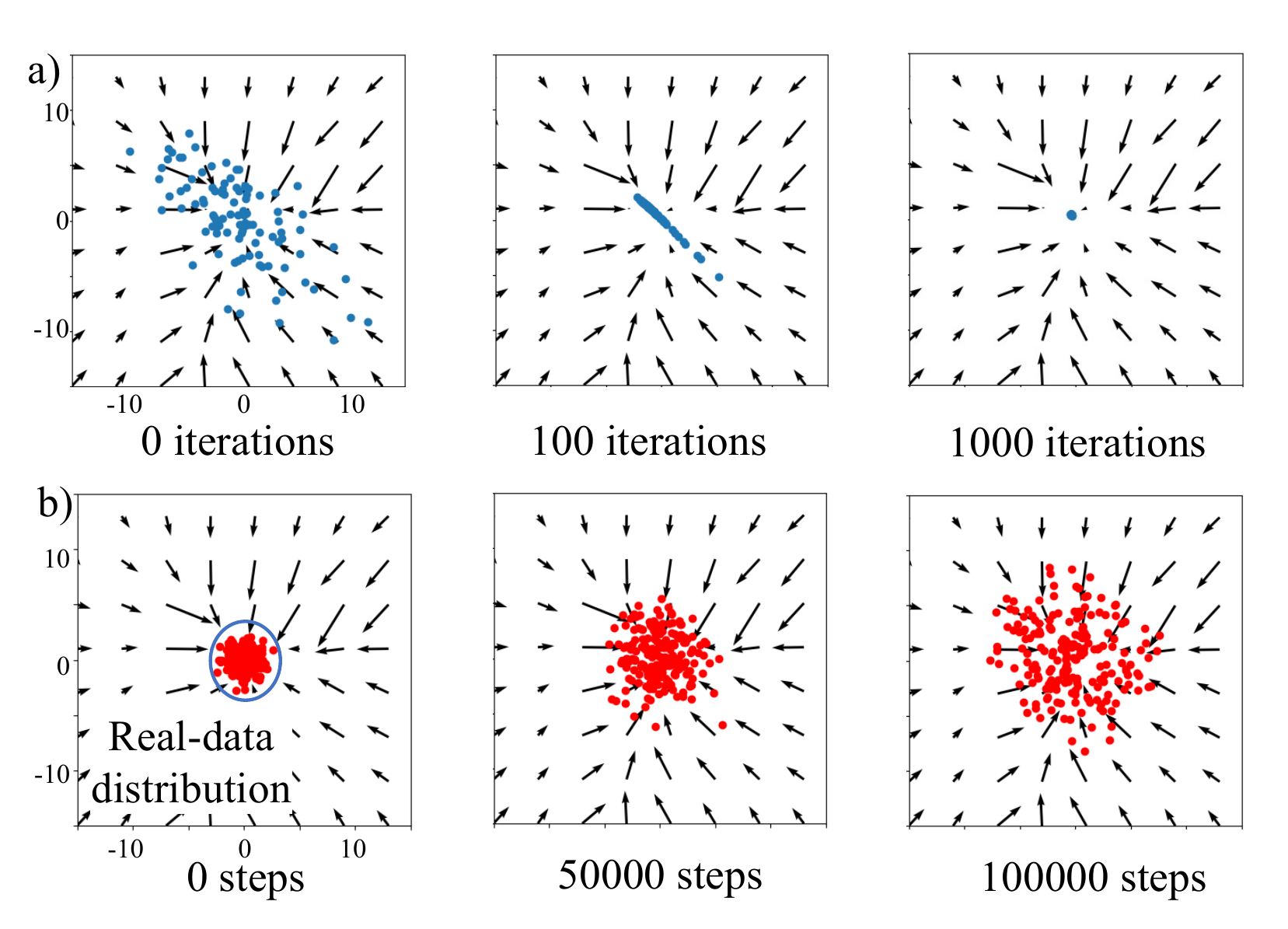}
    {Data generated by a) HumanGAN and b) HumanDiffusion.}
    \drawfig{t}{0.9\linewidth}{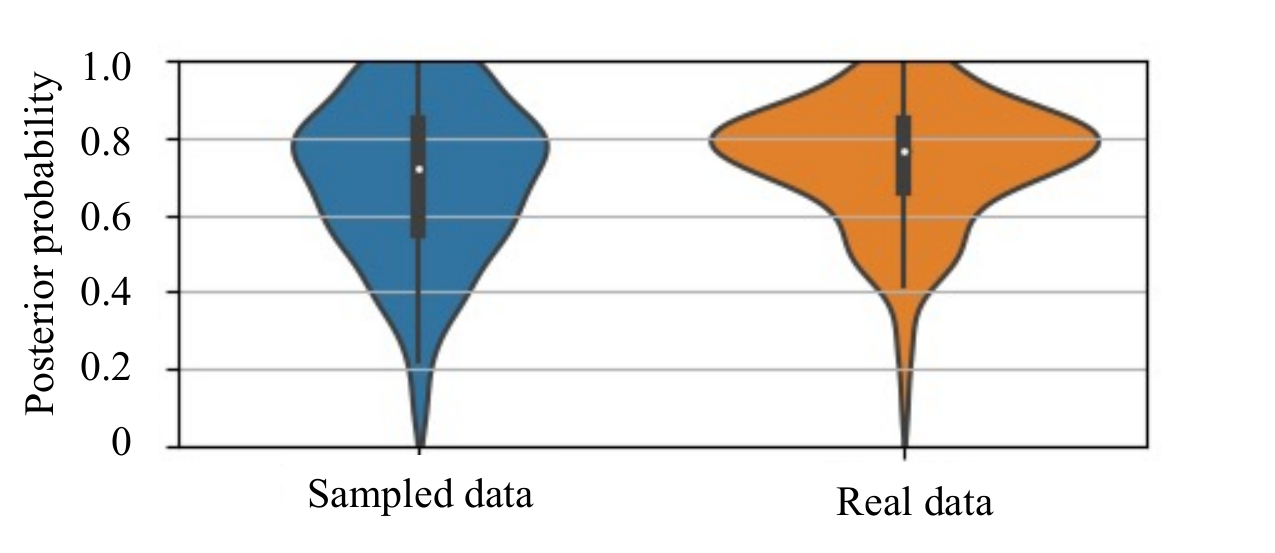}{Violin plots of posterior acceptability. The white point indicates the average.}
    
    \vspace{-1mm}
    \subsection{Evaluation using real and generated data}
    \vspace{-1mm}
        Finally, we quantitatively verified whether HumanDiffusion can generate samples from a human-acceptable distribution wider than a real-data distribution. We prepared two data sets: from the real-data and the HumanDiffusion distributions. The posterior probabilities of these data sets were evaluated using the MUSHRA test. The posterior probability ranged within $[0, 1]$ and corresponds to $D\pt{\x}$. The numbers of the data are 200 respectively. The total number of listeners was $40$. A listener evaluated 20 samples.

        \Fig{violin} shows the violin plots of the posterior probability. The average of real-data distribution data is $73$ and the average of generated data is $69$. Therefore, we consider that the generated data are sufficiently natural, although the variation of the data distribution is much larger than that of the real-data distribution. It is clear that HumanDiffusion can represent a human-acceptable distribution.    
\vspace{-2mm}        
\section{Conclusions}
    \vspace{-2mm}
In this paper, we presented HumanDiffusion, which can represent a human-acceptable (humans' perceptually recognizable) distribution. We evaluated the use of HumanDiffusion in modeling phoneme perception, and we qualitatively and quantitatively demonstrated that HumanDiffusion can represent a human-acceptable distribution. As our future work, we will examine the scalability of the HumanDiffusion in terms of feature dimensionality.

\textbf{Acknowledgements:} This work was spported by JSPS KAKENHI 23H03418, 21H04900, and JST FOREST JPMJFR226V.

\bibliographystyle{IEEEtran}

\bibliography{mybib}

\begin{thebibliography}{10}
\providecommand{\url}[1]{#1}
\csname url@samestyle\endcsname
\providecommand{\newblock}{\relax}
\providecommand{\bibinfo}[2]{#2}
\providecommand{\BIBentrySTDinterwordspacing}{\spaceskip=0pt\relax}
\providecommand{\BIBentryALTinterwordstretchfactor}{4}
\providecommand{\BIBentryALTinterwordspacing}{\spaceskip=\fontdimen2\font plus
\BIBentryALTinterwordstretchfactor\fontdimen3\font minus
  \fontdimen4\font\relax}
\providecommand{\BIBforeignlanguage}[2]{{%
\expandafter\ifx\csname l@#1\endcsname\relax
\typeout{** WARNING: IEEEtran.bst: No hyphenation pattern has been}%
\typeout{** loaded for the language `#1'. Using the pattern for}%
\typeout{** the default language instead.}%
\else
\language=\csname l@#1\endcsname
\fi
#2}}
\providecommand{\BIBdecl}{\relax}
\BIBdecl

\bibitem{goodfellow2020generative}
I.~Goodfellow, J.~Pouget-Abadie, M.~Mirza, B.~Xu, D.~Warde-Farley, S.~Ozair,
  A.~Courville, and Y.~Bengio, ``Generative adversarial networks,''
  \emph{Communications of the ACM}, vol.~63, no.~11, pp. 139--144, 2020.

\bibitem{kingma2013vae}
D.~Kingma and M.~Welling, ``Auto-encoding variational bayes,'' \emph{arXiv},
  vol. abs/1312.6114, 2013.

\bibitem{dinh2014nice}
L.~Dinh, D.~Krueger, and Y.~Bengio, ``{NICE}: Non-linear independent components
  estimation,'' in \emph{Proc. ICLR}, San Diego, U.S.A., 2015.

\bibitem{rombach2022high}
R.~Rombach, A.~Blattmann, D.~Lorenz, P.~Esser, and B.~Ommer, ``High-resolution
  image synthesis with latent diffusion models,'' in \emph{Proceedings of the
  IEEE/CVF Conference on Computer Vision and Pattern Recognition}, 2022, pp.
  10\,684--10\,695.

\bibitem{saito18advss}
Y.~Saito, S.~Takamichi, and H.~Saruwatari, ``Statistical parametric speech
  synthesis incorporating generative adversarial networks,'' \emph{IEEE/ACM
  Transactions on Audio, Speech, and Language Processing}, vol.~26, no.~1, pp.
  84--96, 2018.

\bibitem{li2021starganv2}
Y.~A. Li, A.~Zare, and N.~Mesgarani, ``Starganv2-vc: A diverse, unsupervised,
  non-parallel framework for natural-sounding voice conversion,'' \emph{arXiv
  preprint arXiv:2107.10394}, 2021.

\bibitem{casanova2022yourtts}
E.~Casanova, J.~Weber, C.~D. Shulby, A.~C. Junior, E.~G{\"o}lge, and M.~A.
  Ponti, ``Yourtts: Towards zero-shot multi-speaker tts and zero-shot voice
  conversion for everyone,'' in \emph{International Conference on Machine
  Learning}.\hskip 1em plus 0.5em minus 0.4em\relax PMLR, 2022, pp. 2709--2720.

\bibitem{floridi2020gpt}
L.~Floridi and M.~Chiriatti, ``Gpt-3: Its nature, scope, limits, and
  consequences,'' \emph{Minds and Machines}, vol.~30, pp. 681--694, 2020.

\bibitem{choi2020stargan}
Y.~Choi, Y.~Uh, J.~Yoo, and J.-W. Ha, ``Stargan v2: Diverse image synthesis for
  multiple domains,'' in \emph{Proceedings of the IEEE/CVF Conference on
  Computer Vision and Pattern Recognition}, 2020, pp. 8188--8197.

\bibitem{fujii2020humangan}
K.~Fujii, Y.~Saito, S.~Takamichi, Y.~Baba, and H.~Saruwatari, ``Humangan:
  generative adversarial network with human-based discriminator and its
  evaluation in speech perception modeling,'' in \emph{ICASSP 2020-2020 IEEE
  International Conference on Acoustics, Speech and Signal Processing
  (ICASSP)}.\hskip 1em plus 0.5em minus 0.4em\relax IEEE, 2020, pp. 6239--6243.

\bibitem{zhang2018convergence}
Z.~Zhang, M.~Li, and J.~Yu, ``On the convergence and mode collapse of gan,'' in
  \emph{SIGGRAPH Asia 2018 Technical Briefs}, 2018, pp. 1--4.

\bibitem{song2020score}
Y.~Song, J.~Sohl-Dickstein, D.~P. Kingma, A.~Kumar, S.~Ermon, and B.~Poole,
  ``Score-based generative modeling through stochastic differential
  equations,'' \emph{arXiv preprint arXiv:2011.13456}, 2020.

\bibitem{lambert2022illustrating}
N.~Lambert, L.~Castricato, L.~von Werra, and A.~Havrilla, ``Illustrating
  reinforcement learning from human feedback (rlhf),'' \emph{Hugging Face
  Blog}, 2022, https://huggingface.co/blog/rlhf.

\bibitem{asadi2015human}
H.~Asadi, S.~Mohamed, K.~Nelson, S.~Nahavandi, and D.~R. Zadeh, ``Human
  perception-based washout filtering using genetic algorithm,'' in \emph{Neural
  Information Processing: 22nd International Conference, ICONIP 2015, Istanbul,
  Turkey, November 9-12, 2015, Proceedings, Part II 22}.\hskip 1em plus 0.5em
  minus 0.4em\relax Springer, 2015, pp. 401--411.

\bibitem{ilyas18blackboxlimitedquery}
A.~Ilyas, L.~Engstrom, A.~Athalye, and J.~Lin, ``Black-box adversarial attacks
  with limited queries and information,'' in \emph{Proc. ICML}, vol.~2,
  Stockholm, Sweden, 2018, pp. 2137--2146.

\bibitem{welling2011bayesian}
M.~Welling and Y.~W. Teh, ``Bayesian learning via stochastic gradient langevin
  dynamics,'' in \emph{International Conference on Machine Learning}, 2011, pp.
  681--688.

\bibitem{jvpd_corpus}
``Vowel database: Five {J}apanese vowels of males, females, and children along
  with relevant physical data ({JVPD}),''
  \url{http://research.nii.ac.jp/src/en/JVPD.html}.

\bibitem{morise16world}
M.~Morise, F.~Yokomori, and K.~Ozawa, ``{WORLD}: a vocoder-based high-quality
  speech synthesis system for real-time applications,'' \emph{IEICE
  transactions on information and systems}, vol. E99-D, no.~7, pp. 1877--1884,
  2016.

\bibitem{morise16d4c}
M.~Morise, ``{D4C}, a band-aperiodicity estimator for high-quality speech
  synthesis,'' \emph{Speech Communication}, vol.~84, pp. 57--65, 2016.

\bibitem{lancers}
``Lancers,'' \url{https://www.lancers.jp/}.

\bibitem{liu2016large}
W.~Liu, Y.~Wen, Z.~Yu, and M.~Yang, ``Large-margin softmax loss for
  convolutional neural networks,'' \emph{arXiv preprint arXiv:1612.02295},
  2016.

\bibitem{kingma14adam}
D.~Kingma and B.~Jimmy, ``Adam: A method for stochastic optimization,'' in
  \emph{ar{X}iv preprint ar{X}iv:1412.6980}, 2014.

\end{thebibliography}
\itemsep 30pt 
\end{document}